\documentclass[aps,prd,superscriptaddress,twoside,twocolumn,nofootinbib,showpacs,floatfix]{revtex4-2}
\usepackage{amsmath,amssymb}
\usepackage{graphicx,bm,braket}
\usepackage{subfigure}
\usepackage{booktabs}
\usepackage{dcolumn}
\usepackage{multirow} 
\usepackage{array}
\usepackage{color}
\pdfpagewidth=\paperwidth
\pdfpageheight=\paperheight
\allowdisplaybreaks
\usepackage{epstopdf}
\newcommand*{\slashed}[1]{{#1\!\!\!/}}

\usepackage{float}
\makeatletter

\newcommand{\Rmnum}[1]{\expandafter\@slowromancap\romannumeral #1@}
\makeatother
\usepackage{ulem}

\newcommand{\cmmnt}[1]{}

\usepackage{hyperref}
\usepackage{placeins}
\hypersetup{
colorlinks=true,
linkcolor=red,
citecolor=blue,
urlcolor=black,
}

\begin{document}

\title{\boldmath $\phi$ and $J/\psi$ Production in Proton-Proton Scattering through Hadronic Molecules}

\author{Di Ben}
\email{bd@mail.tsinghua.edu.cn}
\affiliation{Department of Physics and Center for High Energy Physics, Tsinghua University, Beijing 100084, China}

\author{Bing-Song  Zou}
\email{zoubs@mail.tsinghua.edu.cn}
\affiliation{Department of Physics and Center for High Energy Physics, Tsinghua University, Beijing 100084, China}

\date{\today} 

\begin{abstract}
This study estimates the contributions of hidden-strangeness hadronic molecular states ($N(2080)3/2^-$, $N(2270)3/2^-$) to $\phi$ production and hidden-charm $P_c$ states to $J/\psi$ production in $pp$ collisions. The calculated cross-sections reach $\sim 10~\mu b$ for $pp\to pp\phi$ and $\sim 0.1~nb$ for $pp\to pp J/\psi$ at energies $\sim 2$~GeV above threshold. We also compute the spin density matrix element $\rho_{00}$ and the decay angular distributions for $\phi\to K^+K^-$ and $J/\psi\to\mu^+\mu^-$. To support future experiments, we conduct Monte Carlo simulations for the four-body final states of $pp\to ppK^+K^-$ and $pp\to pp\mu^+\mu^-$. The resulting lab-frame distributions are expected to inform detector design at facilities like the High-Intensity heavy-ion Accelerator Facility (HIAF) and the proposed Chinese Advanced Nuclear Physics Research Facility (CNUF), enabling more precise measurements and tighter theoretical constraints. Furthermore, to explain the STAR collaboration's result on $\phi$ meson spin alignment in heavy-ion collisions, we propose a novel mechanism involving hidden-strangeness molecular states, testable with future high-statistics $pp\to pp\phi$ data.

\end{abstract}

 \pacs{}

\keywords{p-p scattering, hadronic molecular, effective Lagrangian approach, nucleon resonances}

\maketitle

\section{Introduction}\label{Sec:intro}

Nucleon-nucleon (NN) scattering serves as a vital experimental tool in particle physics and an essential probe for investigating the non-perturbative regime of Quantum Chromodynamics (QCD).
Higher-energy NN scattering, particularly processes that excite strange and charm quark–antiquark pairs, is of great interest. It offers a new platform for exploring nucleon excitations, thereby significantly advancing our understanding of their spectral and structural properties.

The $\phi$ and $J/\psi$ mesons are pure $s\bar{s}$ and $c\bar{c}$ states, respectively. Studying their production in proton-proton scattering thus provides a crucial avenue for exploring hidden strange and charm quark physics, a topic of significant importance to both experimental and theoretical research. 

With the increase of experimental energy scales, more production thresholds have been opened, and experimentalists have indeed observed processes involved with a pair of strange quarks in proton-proton scattering including $pp \rightarrow pp\phi$, whose cross sections near the threshold have been successively measured by the ANKE Collaboration~\cite{Hartmann:2005wj, Hartmann:2006zc} and the DISTO Collaboration~\cite{DISTO:1998hzu, DISTO:2000dfs}.
To explain the mechanism of $\phi$ meson production in proton-proton scattering—a process expected to be suppressed by the OZI rule—theorists have successfully interpreted the near-threshold experimental data by introducing the exchange of the $N(1535)$ resonance, which itself contains a hidden strange quark pair~\cite{Xie:2007qt}, or without resonance by considering only the nucleonic and mesonic current contribution~\cite{Tsushima:2003fs}.

In future experiments, the being constructed High-Intensity heavy-ion Accelerator Facility (HIAF) and the proposed Chinese Advanced Nuclear Physics Research Facility (CNUF)~\cite{Yang:2013yeb, Zhou:2022pxl, An:2025lws} will be capable of conducting proton-proton scattering experiments with proton beam $E_p = 9.3$ and $25$ GeV, corresponding center-of-mass energies of approximately $4.38$ and $6.98$ GeV, respectively. It is possible to study the higher energy region for $pp\rightarrow pp\phi$ and $pp\rightarrow pp J/\psi$. New experiments may provide crucial data to advance our understanding of the corresponding nucleon excitation spectra.

Furthermore, the study of nucleon structures moving beyond the traditional quark model, including hadronic molecular, compact pentaquarks, and hybrid states, has emerged as a frontier research area in hadron physics.
The discovery of $J/\psi \, p$ resonances in 2015, when the LHCb Collaboration presented striking evidence, named $P_c^+(4380)$ and $P_c^+(4450)$, in $\Lambda^0_b\to K^- J/\psi \, p$ decays~\cite{Aaij:2015tga}. 
Further information was reported in 2019, the LHCb Collaboration declared the $P_c^+(4312)$ state and a two-peak structure of the $P_c^+(4450)$ state, which is resolved into $P_c^+(4440)$ and $P_c^+(4457)$~\cite{Aaij:2019}. These three narrow hidden-charm states are located close to the $\Sigma_c \bar D^{(*)}$ thresholds, supporting earlier predictions~\cite{Wu:2010jy,Wu:2010vk,Wang:2011rga,Wu:2012md}  in the picture of hadronic molecules.

Theoretical studies of hadronic molecular states with a hidden-charm quark pair have been extended, and their detailed decay properties have been predicted \cite{Lin:2017mtz, Shen:2017oyw, Lin:2019qiv}. Recently, the study of effects of $P_c$ states in photoproduction $\gamma p\rightarrow J/\psi p$~\cite{Duan:2024hby}, which introduced $P_c(4312)$, $P_c(4380)$, $P_c(4440)$, and $P_c(4457)$ in s-channel, shows that the molecular picture is consistent with the existing experimental data. Considering a huge enhancement raised from the $\bar{D}^\ast\Sigma_c^\ast$ molecule $P_c(4500)1/2^-$ in the precesses in $J/\psi p$~\cite{Lin:2019qiv}, and non-negligible contribution from the $\bar{D}^\ast\Sigma_c^\ast$ molecule $P_c(4511)3/2^-$, we introduce above six $P_c$ states in the estimation of the production of $J/\psi$ mesons.

Similarly, systems containing a hidden-charm quark pair can be analogized to those with a hidden-strangeness pair, allowing us to propose a series of partners to the $P_c$ states—denoted here as $K\Sigma^\ast$, $K^\ast\Sigma$, and $K^\ast\Sigma^\ast$ bound states, namely $N(1875)$, $N(2080)$ and $N(2270)$ respectively—in the strange quark sector. The decay properties of these states have also been systematically investigated~\cite{Lin:2018kcc, Ben:2024qeg}.
In recent theoretical analyses of photoproduction $\gamma p\rightarrow p\phi$, distinct structures have been observed around energies of $2080$ and $2270$ MeV. The experimental data in this process can be well explained by contributions from hadronic molecular states $N(2080)3/2^-$ and $N(2270)3/2^-$ at these energies, which play an essential role in the production of $\phi$ mesons~\cite{Wu:2023ywu}.

In this article, we introduce corresponding hadronic molecular states $N(2080)3/2^-$ and $N(2270)3/2^-$ for the $\phi$ production process $pp\rightarrow pp\phi$, while six $P_c$ states are introduced for $J/\psi$ production process  $pp\rightarrow pp J/\psi$, and calculate their respective production cross sections. In addition, we provide detailed predictions for the spin observables of the final-state $\phi$ and $J/\psi$ mesons at center of mass energy $4.38$ and $6.98$ GeV, respectively, and analyze the influence of intermediate states with different spin parity on the spin density matrix element $\rho_{00}$ leading to a new mechanism proposed for the large global spin alignment of $\phi$ meson. Further calculation of the distribution of the cascade decay of $\phi\rightarrow K^+K^-$ and $J/\psi\rightarrow\mu^+\mu^-$ in its rest frame is also provided. Finally, from the perspective of experiments, we employ Monte Carlo simulations for four-body final states in $pp\rightarrow ppK^+K^-$ and $pp\rightarrow pp\mu^+\mu^-$ to calculate the momentum and angular distributions.

This paper is organized as follows. In Sec. \ref{Sec:forma}, we briefly introduce the framework of our theoretical model. In Sec. \ref{Sec:results}, the results of our theoretical calculations with some discussions are presented. Finally, in Sec. \ref{sec:summary}, we give a brief summary and conclusions.

\section{Formalism}\label{Sec:forma}

\begin{figure}[H]
    \centering
    {\vglue 0.15cm}
    \subfigure[~$\phi$ production]{
    \includegraphics[width=0.36\textwidth]{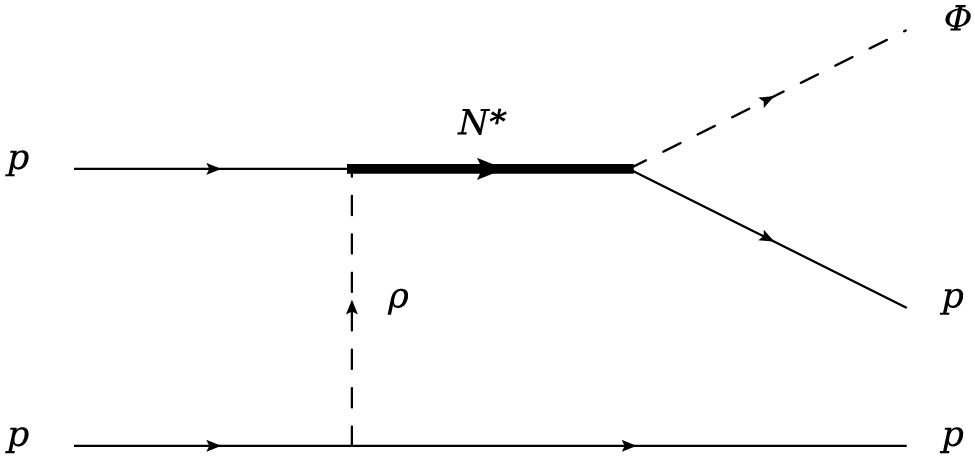}}
    \\[6pt]
    \subfigure[~$J/\psi$ production]{
    \includegraphics[width=0.36\textwidth]{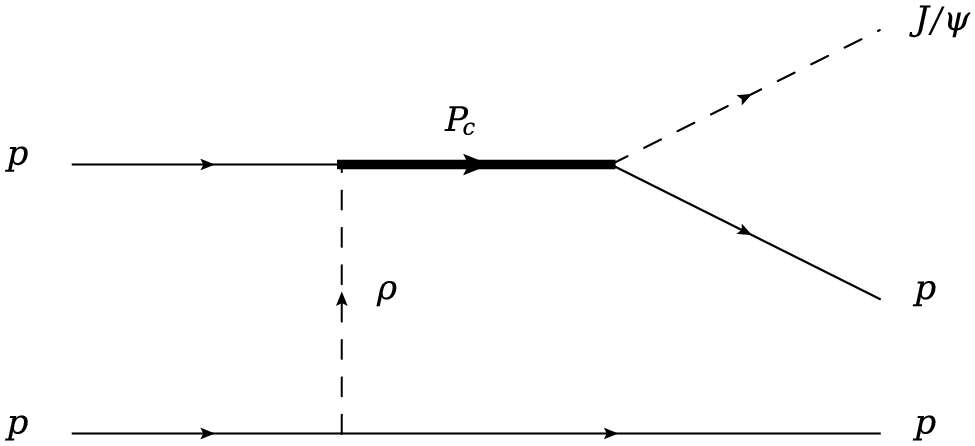}}
    \caption{Generic structures of the $\phi$ and $J/\psi$ production in proton-proton scattering within hadronic molecular picture. Time proceeds from left to right.}
    \label{FD}
\end{figure}

To estimate the production cross sections for $\phi$ and $J/\psi$, we introduce the generic structures for the processes $pp\rightarrow pp\phi$ and $pp\rightarrow pp J/\psi$ within the hadronic molecular picture, as illustrated in Fig.~\ref{FD}.

Given that the emitted mesons in the final state, $\phi$ and $J/\psi$, are both vector mesons, we employ a mechanism involving the exchange of a vector meson $\rho$ to describe the proton-proton interaction.
The production of both $\phi$ and $J/\psi$ involves a common vertex, namely the $NN\rho$ vertex. Its Lagrangian is given by:
\begin{equation}
    \mathcal{L}_{NN\rho} = -g_{NN\rho}\Bar{N}(\gamma_\mu-\frac{\kappa_\rho}{2M_N}\sigma_{\mu\nu}\partial^\nu)\rho^\mu N,\label{lg_rho}
\end{equation}
where the $M_N$ indicates the mass of proton, and the coupling constant $g_{NN\rho} = 3.25$ and $\kappa_\rho = 6.1$, respectively\cite{Koch:1985bp,Ronchen:2012eg}.

\subsection{$pp\rightarrow pp\phi$}\label{sec_2a}

Based on the analysis of molecular states in the photoproduction reaction $\gamma p \rightarrow \phi p$~\cite{Wu:2023ywu}, it was shown that the hidden-strange molecular states $N(2080)3/2^-$ and $N(2270)3/2^-$ dominate the $s$-channel contribution. Consequently, in this work, we employ these same states and their fitted widths from Ref.~\cite{Wu:2023ywu} to estimate the $\phi$ meson production cross section for the process $pp \rightarrow pp\phi$, as listed in Tab.~\ref{tab_N}.

 \begin{table}[H]
    \caption{$N^\ast$ states}
    \belowrulesep=0pt
\aboverulesep=0pt
\renewcommand{\arraystretch}{1.5}
    \centering
    \begin{tabular}{c|c|c|c|c}
       State & Molecule & $J^P$\cite{Wu:2023ywu} & $M_{N^\ast}$(MeV) & $\Gamma_{N^\ast}$(MeV)\cite{Wu:2023ywu} \\\hline
         $N(2080)$ & $K^\ast\Sigma$ & $3/2^-$ & $2080$ & $100$ \\
         $N(2270)$ & $K^\ast\Sigma^\ast$ & $3/2^-$ & $2270$ & $230$ \\\hline
    \end{tabular}
    \label{tab_N}
\end{table}
To maintain consistency with the parameter relations obtained from previous fits\cite{Wu:2023ywu}, we employ the same effective Lagrangian as used in prior work to describe the $RNV$ interaction:
\begin{align}
    \mathcal{L}^{3/2^-}_{RNV} = &-i\frac{g_1}{2M_N}\Bar{R}_\mu\gamma_\nu V^{\mu\nu}N+\frac{g_2}{(2M_N)^2}\Bar{R}_\mu V^{\mu\nu} \partial_\nu N \nonumber\\
    &+ \frac{g_3}{(2M_N)^2}\Bar{R}_\mu (\partial_\nu V^{\mu\nu}) N + h.c.,\label{eq_3half}
\end{align}
where:
\begin{equation}
    V^{\mu\nu} = \partial^\mu V^\nu -\partial^\nu V^\mu
\end{equation}

In decay calculations within the hadronic molecular picture, the branching ratios of molecular states show less sensitivity to the choice of cutoff parameters than the computed decay widths~\cite{Ben:2024qeg}. Hence, the branching ratio (BR) constitutes a more reliable physical observable. Adopting the branching ratios from Refs.~\cite{Lin:2018kcc} and~\cite{Ben:2024qeg} as benchmarks and incorporating the proportional relations among the coupling constants $g_1$, $g_2$, and $g_3$ fitted in Ref.~\cite{Wu:2023ywu}, we determine the coupling constants for the vertex describing the coupling between molecular states and $\phi p$, as listed in Tab.~\ref{tab_couplings}.

 \begin{table}[H]
    \caption{Couping constants of $N^\ast$}
    \belowrulesep=0pt
\aboverulesep=0pt
\renewcommand{\arraystretch}{1.5}
    \centering
    \begin{tabular}{c|c|c|c|c}
       & Vertex & ~~~$g_1$~~~ & ~~~$g_2$~~~ & ~~~$g_3$~~~  \\\hline
        Determined & $N(2080)\rightarrow\phi p$ & $0.50$ & $1.21$ & $2.46$ \\
        & $N(2270)\rightarrow\phi p$ & $0.42$ & $-0.34$ & $0.42$ \\\hline
       Solution A & $N(2080)\rightarrow\rho p$ & $0.21$ & $-0.20$ & $0.21$ \\
        & $N(2270)\rightarrow\rho p$ & $0.30$ & $-0.28$ & $0.30$ \\\hline
        Solution B & $N(2080)\rightarrow\rho p$ & $0.15$ & $-0.31$ & $0.13$ \\
        & $N(2270)\rightarrow\rho p$ & $0.20$ & $-0.40$ & $0.17$ 
    \end{tabular}
    \label{tab_couplings}
\end{table}

Indeed, a degree of freedom remains in the coupling to $\rho p$. For simplicity, we assume that $N(2080)3/2^-$ shares the same coupling structure with $N(2270)3/2^-$. Thanks to the availability of experimental data near the production threshold, the relative coupling strength between $N(2080)3/2^-$ and $N(2270)3/2^-$ to $\rho p$ can be constrained. This allows our model to not only describe the $\gamma p \rightarrow \phi p$ photoproduction satisfactorily but also partially reproduce the cross-section behavior of $pp\rightarrow pp\phi$ scattering near the threshold. We find two characteristic solutions, labeled Solution A and Solution B, whose corresponding coupling constants are also listed in Tab.~\ref{tab_couplings}.

Taking $N(2270)\rightarrow\phi p$ as an example, the partial decay width is given by $\Gamma_{N(2270)\rightarrow\phi p} = \Gamma_{N(2270)} \times \text{BR}_{N(2270)\rightarrow\phi p}$. Adopting the branching ratio $\text{BR}_{N(2270)\rightarrow\phi p} \sim 5\%$ from Ref.~\cite{Ben:2024qeg} yields $\Gamma_{N(2270)\rightarrow\phi p} = 11.6$ MeV. Furthermore, the proportional relations among the coupling constants $g_1$, $g_2$, and $g_3$ from Ref.~\cite{Wu:2023ywu} are $g_1:g_2:g_3 = 1:-0.8:1$. Therefore, by calculating the decay process $N(2270)\rightarrow\phi p$, the corresponding coupling constants can be determined.

\subsection{$pp\rightarrow pp J/\psi$}

In a previous study of $\gamma p \rightarrow J/\psi p$~\cite{Duan:2024hby}, four $P_c$ states---$P_c(4312)$, $P_c(4380)$, $P_c(4440)$, and $P_c(4457)$---were introduced. The deficit observed in the differential cross sections at higher energies indicates that additional $s$-channel processes may contribute. Calculations of decay patterns for $P_c$ states reveal a significant enhancement from the $\bar{D}^\ast\Sigma_c^\ast$ molecule $P_c(4500)1/2^-$, along with a non-negligible contribution from the $\bar{D}^\ast\Sigma_c^\ast$ molecule $P_c(4511)3/2^-$ in the $J/\psi p$ channel~\cite{Lin:2019qiv}. These two states, which provide notable contributions to $J/\psi$ production, coincidentally match the unexplained higher-energy structure reported in Ref.~\cite{Duan:2024hby}. Therefore, in the present work on the proton-proton scattering process $pp \rightarrow ppJ/\psi$, we introduce a total of six corresponding $P_c$ states. Their assigned spin-parity quantum numbers, masses, and decay widths are listed in Tab.~\ref{tab_Pc}.
\begin{table}[H]
    \caption{$P_c$ states}
    \belowrulesep=0pt
\aboverulesep=0pt
\renewcommand{\arraystretch}{1.5}
    \centering
    \begin{tabular}{c|c|c|c|c}
       State & Molecule & $J^P$\cite{Xie:2022hhv} & $M_{P_c}$(MeV) & $\Gamma_{P_c}$(MeV) \\\hline
         $P_c(4312)$ & $\Bar{D}\Sigma_c$ & $1/2^-$ & $4311.9 \cite{ParticleDataGroup:2024cfk}$ & $9.8 \cite{ParticleDataGroup:2024cfk}$ \\
         $P_c(4380)$ & $\Bar{D}\Sigma_c^\ast$ & $3/2^-$ & $4372.2$ \cite{Xie:2022hhv}& $9.6$ \cite{Xie:2022hhv}\\
         $P_c(4440)$ & $\Bar{D}^\ast\Sigma_c$ & $1/2^-$ & $4440.3 \cite{ParticleDataGroup:2024cfk}$ & $20.6 \cite{ParticleDataGroup:2024cfk}$ \\
         $P_c(4457)$ & $\Bar{D}^\ast\Sigma_c$ & $3/2^-$ & $4457.3 \cite{ParticleDataGroup:2024cfk}$ & $6.4 \cite{ParticleDataGroup:2024cfk}$ \\
         $P_c(4500)$ & $\Bar{D}^\ast\Sigma_c^\ast$ & $1/2^-$ & $4502.7$ \cite{Xie:2022hhv}& $28$ \cite{Xie:2022hhv}\\
         $P_c(4511)$ & $\Bar{D}^\ast\Sigma_c^\ast$ & $3/2^-$ & $4510.5$ \cite{Xie:2022hhv}& $14.4$\cite{Xie:2022hhv} \\\hline
    \end{tabular}
    \label{tab_Pc}
\end{table}

We consider two types of $P_c$ states with quantum numbers $J^P=1/2^-$ and $3/2^-$, respectively. Following Ref.~\cite{Duan:2024hby}, we employ the same effective Lagrangian to describe the coupling between the $P_c$ states, the vector meson, and the proton. The explicit forms are given by:
\begin{align}
    \mathcal{L}_{P_c(1/2^-)V p} = & g_{P_c(1/2^-)}\bar{N}\gamma_5\gamma_\rho\left(-g^{\rho\mu}+\frac{p^\rho p^\mu}{M_{P_c}^2}\right)P_c V_\mu, \\
    \mathcal{L}_{P_c(3/2^-)V p} = & g_{P_c(3/2^-)}\bar{N}P_c^\mu V_\mu,
\end{align}
where $p$ denotes the 4-momentum of the $P_c$ states, and $V$ represents either $J/\psi$ or $\rho$.

Based on the analysis of $P_c$ states in the photoproduction reaction $\gamma p \rightarrow J/\psi p$~\cite{Duan:2024hby}, it is known that although hidden-charm hadronic molecular $P_c$ states are consistent with experimental data, their coupling constants $g_c$ should not take large values. Therefore, in this work we adopt $g_c(V=J/\psi)=0.05$ and $g_c(V=\rho)=0.01$ for the respective vertices.

\subsection{Propagators}
For both $\phi$ and $J/\psi$ production process, the propagator of spin $3/2$ particles need to be introduced:
\begin{equation}
    S_{3/2}(q) = \frac{i}{\slashed q-M_R+i\Gamma_R/2}\left(\Tilde{g}_{\mu\nu} + \frac{1}{3}\Tilde{\gamma}_\mu\Tilde{\gamma}_\nu\right),
\end{equation}
where q is the 4-momentum of the exchange $N^*$ or $P_c$ state, $M_R$ and $\Gamma_R$ is the corresponding mass and width, with:
\begin{align}
    \Tilde{g}_{\mu\nu} &= -g_{\mu\nu} + \frac{q_\mu q_\nu}{M_R^2},\\
    \Tilde{\gamma} =& \gamma^\nu \Tilde{g}_{\mu\nu} = -\gamma_\mu + \frac{q_\mu\slashed q}{M^2_R}
\end{align}

For the $\phi$ meson production process, we introduce the conventional propagator form for the intermediate spin-1 particle $\rho$:
\begin{equation}
D_{\mu\nu}(q) = \frac{-i(g_{\mu\nu} - q_\mu q_\nu / M_\rho^2)}{q^2 - M_\rho^2 + i\epsilon}
\label{eq:phi_propagator}
\end{equation}

For the $J/\psi$ production process, due to the high mass scale of the intermediate $P_c$ state, we employ a Reggeized propagator:
\begin{equation}
\mathcal{D}_{\mu\nu}(s,t) = D_{\mu\nu}(t) \cdot \left(\frac{s}{s_0}\right)^{\alpha(t)-1} \frac{\pi\alpha'}{\sin[\pi\alpha(t)]} \frac{\mathcal{S}+e^{-i\pi\alpha(t)}}{2},
\label{eq:regge_propagator}
\end{equation}
where:
\begin{equation}
    D_{\mu\nu}(t) = \frac{-i(g_{\mu\nu} - q_\mu q_\nu / M_\rho^2)}{\Gamma(\alpha(t))}.
\end{equation}
And the Regge trajectory of $\rho$ meson is given by \cite{Guidal:1997hy}:
\begin{equation}
\alpha(t) = \alpha_0 + \alpha' t = 0.55+0.8t
\label{eq:regge_trajectory}
\end{equation}
Here $\mathcal{S}$ represents the signature factor, which for $\rho$ mesons is $\mathcal{S} = +1$.

\subsection{Form Factors}

To regularize the high-energy behavior of the amplitude, we introduce two form factors in the calculation, corresponding to $\rho$ meson exchange and excited nucleon exchange, respectively. 
For the $\rho$ meson exchange, the form factor is written as:
\begin{equation}
    f_\rho(l^2) = \left(\frac{\Lambda_\rho^2-M_\rho^2}{\Lambda_\rho^2-l^2}\right)^2,
\end{equation}
where $l$ is the 4-momentum of the exchange $\rho$ meson. The value of the cutoff parameter $\Lambda_\rho$ is empirically chosen as $M_\rho+500$ MeV. Thus, we take $\Lambda_\rho=1300$ MeV.

For the $N^\ast$ or $P_c$ states, the form factor is written as:
\begin{equation}
    f_N(q^2) = \left(\frac{\Lambda_N^4}{\Lambda^4_N+(q^2-M_N^2)^2}\right)^2,
\end{equation}
where $q$ is the 4-momentum of the exchange $N^\ast$ or $P_c$ states. The value of the cutoff parameter $\Lambda_N$ is taken from Ref.\cite{Wu:2023ywu}, which is $\Lambda_N = 1500$ MeV.
\subsection{Spin Density Matrix}

To characterize the spin properties of the final-state vector mesons, we introduce the spin density matrix~\cite{Titov:1997kt}.
In the helicity basis, the amplitude can be written as:
\begin{equation}
T_{\lambda_V \lambda}(W, \theta)
\end{equation}
where $W$ is the center-of-mass energy, $\theta$ is the scattering angle, and $\lambda_V$ represent the helicity of the outgoing vector particle, and $\lambda$ represent the helicities of all other particles. This enables the definition of the spin density matrix $\rho$ at a fixed center-of-mass energy $W$ as a function of the scattering angle $\theta$:
\begin{equation}
\rho_{\lambda_V\lambda_V'}(W,\theta) = \frac{\sum_{\lambda} T_{\lambda_V \lambda}(W,\theta) T_{\lambda_V' \lambda}^*(W,\theta)}{\sum_{\lambda_V\lambda}|T_{\lambda_V\lambda}(W,\theta)|^2}  .
\end{equation}

Based on these properties, we can further predict the angular distribution of its cascade decay products in the rest frame of the vector meson~\cite{Titov:1997kt}.

For $\phi$ meson, we can further calculate the distribution $\mathcal{W}^{K^+K^-}(cos\Theta)$ of $K^+K^-$ particles in the rest frame of $\phi$ meson, where $\Theta$ is the decay angle, which is:
\begin{equation}
    \mathcal{W}^{K^+K^-}(cos\Theta) = \frac{3}{4}(1-\rho_{00})(1+B^{K^+K^-}\cos^2\Theta),
\end{equation}
where $B^{K^+K^-}$ is called decay anisotropy given by:
\begin{equation}
    B^{K^+K^-} = -\frac{1-3\rho_{00}}{1-\rho_{00}}.
\end{equation}

For $J/\psi$ meson, we can further calculate the distribution $\mathcal{W}^{\mu^+\mu^-}(cos\Theta)$ of $\mu^+\mu^-$ particles in the rest frame of $J/\psi$ meson, where $\Theta$ is the decay angle, which is:
\begin{equation}
    \mathcal{W}^{\mu^+\mu^-}(cos\Theta) = \frac{3}{8}(1+\rho_{00})(1+B^{\mu^+\mu^-}\cos^2\Theta),
\end{equation}
where $B^{\mu^+\mu^-}$ is given by:
\begin{equation}
    B^{\mu^+\mu^-} = \frac{1-3\rho_{00}}{1+\rho_{00}}.
\end{equation}


\section{Results and Discussion}\label{Sec:results}

\subsection{$pp\rightarrow pp\phi$}

\subsubsection{Cross Section}

\begin{figure}[htb]
    \vglue 6pt
    \centering
    \includegraphics[width=0.48\textwidth]{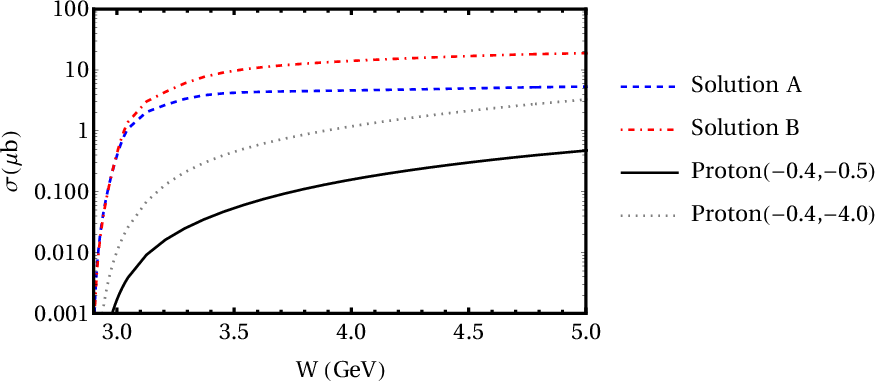}
    \caption{Total cross sections obtained by our model and simply estimation of the contribution of the proton, where the blue dashed line and red dot-dashed line corresponding to the solution A and B listed in Tab.~\ref{tab_couplings}, while the black solid line and the gray dotted line corresponding to the $(g_{\phi NN},\kappa_{\phi}) = (-0.4,-0.5)$ and $(g_{\phi NN},\kappa_{\phi}) = (-0.4,-4.0)$ case based on Ref.~\cite{Tsushima:2003fs}.}
    \label{Total1}
\end{figure}

\begin{figure}[htb]
    \vglue 6pt
    \centering
    \includegraphics[width=0.44\textwidth]{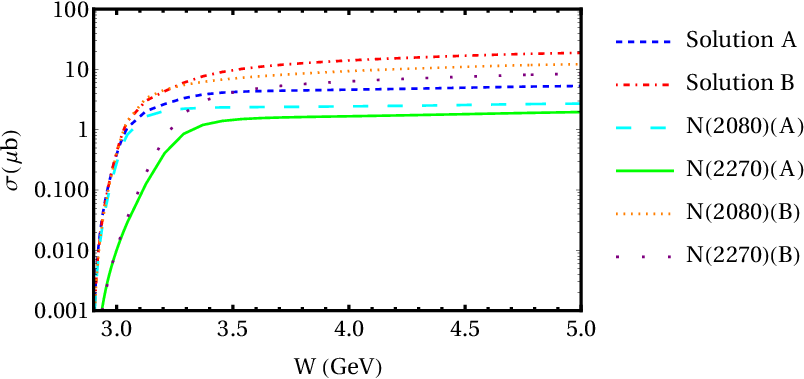}
    \caption{Total cross sections and the single molecular state $N(2080)3/2^-$ or $N(2270)3/2^-$ contribution to the cross section. Where the blue dashed line and red dot-dashed line notably the same as its in Fig.\ref{Total1}, while the cyan large-dashed line, green solid line, orange dotted line and purple large-dotted line corresponding to the contribution from $N(2080)3/2^-$ in solution A, $N(2270)3/2^-$ in solution A, $N(2080)3/2^-$ in solution B, and $N(2270)3/2^-$ in solution B, respectively. Solutions A and B are based on Tab.~\ref{tab_couplings}.}
    \label{Total2}
\end{figure}

Figs.~\ref{Total1} and~\ref{Total2} present the cross sections predicted by our model, as introduced in Sec.~\ref{sec_2a}. The model utilizes existing photoproduction data to constrain the relevant coupling constants, enabling predictions for $pp\rightarrow pp\phi$ up to a center-of-mass energy of $W=5$ GeV. This energy range covers the region of interest at $W=4.38$ GeV, corresponding to a proton beam energy of $9.3$ GeV. Our obtained Solutions A and B are plotted as the blue dashed and red dot-dashed lines, respectively, in these figures.

We also consider the possibility that the proton may possess an intrinsic strange quark component. Following Ref.~\cite{Tsushima:2003fs}, the proton-$\phi$ meson coupling is well-defined, with the Lagrangian identical to Eq.~\eqref{lg_rho} upon replacing $\rho$ with $\phi$. The corresponding form factors and cutoff parameters are also determined, allowing a straightforward description of the proton contribution based on a simple $\rho$-exchange mechanism. We select the parameter sets $(g_{\phi NN},\kappa_{\phi}) = (-0.4,-0.5)$ and $(g_{\phi NN},\kappa_{\phi}) = (-0.4,-4.0)$, which were compared in Ref.~\cite{Hartmann:2005wj}, to estimate the proton contribution. These contributions are shown in Fig.~\ref{Total1}, with the black solid line corresponding to $(-0.4,-0.5)$ and the gray dotted line to $(-0.4,-4.0)$.

As shown by the blue dashed and red dot-dashed lines in Fig.~\ref{Total1}, Solutions A and B exhibit distinct behaviors above a center-of-mass energy of $W \sim 3$ GeV. Solution A flattens out, with its cross section stabilizing around $5~\mu$b, while Solution B continues to rise gradually, eventually reaching $20~\mu$b. This difference can be attributed to the specific ratio between the coupling constants $g_2$ and $g_3$ listed in Tab.~\ref{tab_couplings}. Overall, the total cross section for $\phi$ meson production in $pp\rightarrow pp\phi$ scattering is on the order of $10~\mu$b.

The black solid and gray dotted lines in Fig.~\ref{Total1} correspond to the proton intrinsic strangeness contribution with parameter sets $(g_{\phi NN},\kappa_{\phi}) = (-0.4,-0.5)$ and $(-0.4,-4.0)$, respectively, based on Ref.~\cite{Tsushima:2003fs}. According to Refs.~\cite{Tsushima:2003fs, Hartmann:2005wj}, while $g_{\phi NN}$ can be well determined from near-threshold data, the tensor coupling $\kappa_\phi$ remains poorly constrained within a large range. Although both curves exhibit similar shapes, a larger tensor coupling leads to a significantly enhanced cross section: at $W = 5$ GeV, $\kappa_\phi = -0.5$ yields $0.5~\mu$b, while $\kappa_\phi = -4.0$ gives $3~\mu$b. Thus, the proton contribution to $pp\rightarrow pp\phi$ is approximately one order of magnitude smaller than that from the molecular states.

As shown in Fig.~\ref{Total2}, the cyan large-dashed, green solid, orange dotted, and purple large-dotted lines correspond to the contributions from $N(2080)3/2^-$ in Solution A, $N(2270)3/2^-$ in Solution A, $N(2080)3/2^-$ in Solution B, and $N(2270)3/2^-$ in Solution B, respectively, with both solutions based on the parameters in Tab.~\ref{tab_couplings}.

Although Solutions A and B exhibit the distinct overall trends mentioned above, the individual contributions from each molecular state follow a similar pattern in both solutions. The curves from Solution A are much flatter than those from Solution B. In both cases, $N(2080)3/2^-$ dominates at lower energies (below $W \sim 3.5$ GeV), while $N(2270)3/2^-$ rises to comparable magnitude at higher energies. Specifically, in Solution A the contributions reach $3~\mu$b for $N(2080)3/2^-$ and $2~\mu$b for $N(2270)3/2^-$, whereas in Solution B they reach $10~\mu$b and $7~\mu$b, respectively.

\begin{figure}[H]
    \vglue 6pt
    \centering
    \includegraphics[width=0.35\textwidth]{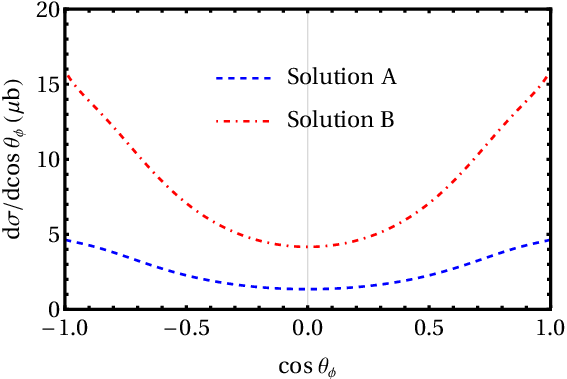}
    \caption{Differential cross section with the angle of outgoing $\phi$ meson $\theta_\phi$ in the center-of-mass frame at the center-of-mass energy $W=4.38$ GeV. Where the blue dashed line and the red dot-dashed line correspond to the solutions A and B obtained by our model.}
    \label{fig_dcs}
\end{figure}

As a prediction at the center-of-mass energy $W=4.38$ GeV, corresponding to a proton beam energy of $E_p = 9.3$ GeV, we present the differential cross section as a function of the outgoing $\phi$ meson angle in the center-of-mass frame. As shown in Fig.~\ref{fig_dcs}, both Solutions A and B exhibit the same characteristic pattern: a dip in the central region and rises at both ends.

To further explore the molecular spectrum and account for possible higher-spin particle production in proton-proton collisions, we introduce the $N(2270)5/2^-$ molecular state predicted in the literature~\cite{Ben:2024qeg}, and qualitatively discuss the properties of such high-spin resonances.

From the perspective of spin-orbital coupling, the crucial difference between $J^P = 3/2^-$ and $J^P = 5/2^-$ states coupled to a vector meson and a baryon ($VB$ system) lies in the allowed orbital angular momentum $L$. A $J^P = 3/2^-$ state can couple via $s$-wave, whereas a $J^P = 5/2^-$ state cannot.

The Lagrangian of a $J^P = 5/2^-$ particle coupled with $VB$ system can be written as:
\begin{align}
    \mathcal{L}^{5/2^-}_{RNV} = &+\frac{g_1}{(2M_N)^2}\Bar{R}_{\mu\alpha}\gamma_\nu\gamma_5 (\partial^\alpha V^{\mu\nu}) N \nonumber\\
    &-i\frac{g_2}{(2M_N)^3}\Bar{R}_{\mu\alpha}\gamma_5 (\partial^\alpha V^{\mu\nu} )\partial_\nu N \nonumber\\
    &+i \frac{g_3}{(2M_N)^3}\Bar{R}_{\mu\alpha}\gamma_5 (\partial^\alpha\partial_\nu V^{\mu\nu}) N + h.c..
    \label{eq_5h}
\end{align}
Compared to Eq.~\eqref{eq_3half}, the key distinction lies in the higher order of partial derivatives applied to the vector field in the Lagrangian. As the center-of-mass energy increases, the momentum of the exchanged vector particle grows, which significantly enhances the resulting cross section.

\begin{figure}[H]
    \vglue 6pt
    \centering
    \includegraphics[width=0.42\textwidth]{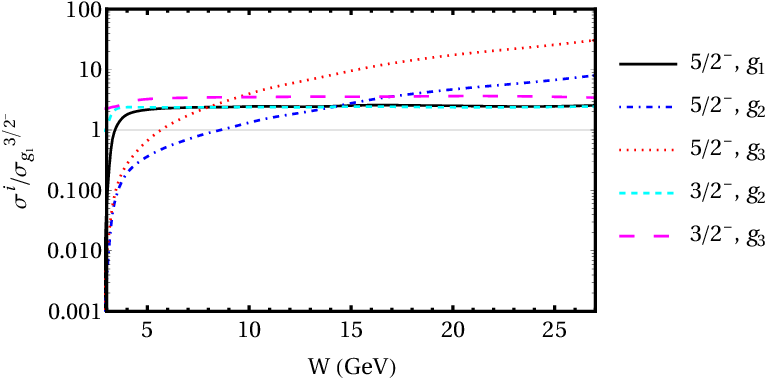}
    \caption{Contribution from different terms individually while the assumption decay widths are equal, re-normalized to $\sigma^{3/2^-}_{g_1}$. Where the gray baseline, cyan dashed line, magenta large-dashed line, black solid line, blue dot-dashed line, and red dotted line correspond to the contribution from $g_1$, $g_2$, and $g_3$ of $J^P = 3/2^-$ and $5/2^-$, respectively.}
    \label{fig_higher_spin}
\end{figure}

Since the relative coupling strengths at the vertices are unknown, we calculate and compare all six terms individually. Under the assumption that the decay widths for $N(2270)\rightarrow\phi N$ and $N(2270)\rightarrow\rho N$ are equal, the resulting cross sections are shown in Fig.~\ref{fig_higher_spin}. As the absolute cross section magnitude is not physically meaningful in this context, we normalize all contributions to that of the $g_1$ term with $J^P = 3/2^-$, denoted as $\sigma^{3/2^-}_{g_1}$. Consequently, the $y$-axis represents the ratio of each contribution relative to $\sigma^{3/2^-}_{g_1}$, while the $x$-axis shows the center-of-mass energy $W$.

As shown in Fig.\ref{fig_higher_spin}, the gray grid line, cyan dashed line, magenta large-dashed line, black solid line, blue dot-dashed line, and red dotted line correspond to the contributions from the $g_1$, $g_2$, and $g_3$ terms for $J^P = 3/2^-$ and $5/2^-$, respectively. While the $g_1$ term for $J^P=5/2^-$ flattens at high energies—similar to the $g_2$ and $g_3$ terms for $J^P=3/2^-$—the $g_2$ and $g_3$ terms for $J^P=5/2^-$ continue to rise with increasing center-of-mass energy. Moreover, at $W = 10$ GeV and $15$ GeV, the contributions from the $g_2$ and $g_3$ terms of $J^P=5/2^-$ surpass those from all other terms.

\begin{figure}[H]
    \centering
    \includegraphics[width=0.42\textwidth]{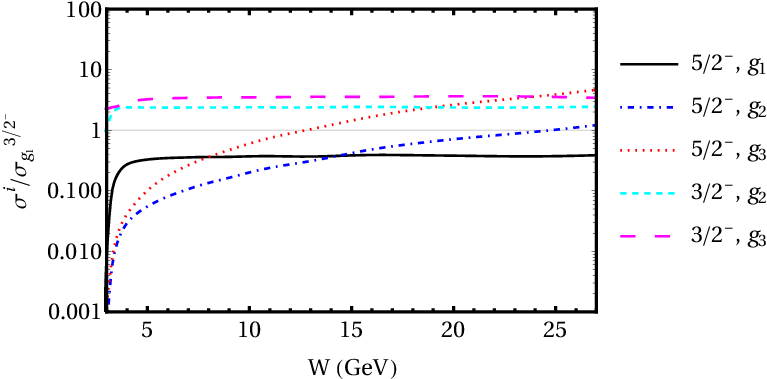}
    \vglue 6pt
    \caption{Contribution from different terms individually while the assumption decay widths are following Ref.\cite{Ben:2024qeg}, re-normalized to $\sigma^{3/2^-}_{g_1}$. Notation is the same as Fig.\ref{fig_higher_spin}.}
    \label{fig_higher_spin2}
\end{figure}

If we adopt the decay widths and branching ratios from Ref.\cite{Ben:2024qeg}, where the width values for $J^P=3/2^-$ and $5/2^-$ states differ within the hadronic molecular picture, the resulting cross sections, which are also normalized to $\sigma^{3/2^-}_{g_1}$, are shown in Fig.\ref{fig_higher_spin2}.

As shown in Fig.\ref{fig_higher_spin2}, even though the coupling for the $J^P=5/2^-$ state is weaker than that for the $J^P=3/2^-$ state according to previous theoretical results~\cite{Ben:2024qeg}, the rising trend of the cross section persists. Within the effective Lagrangian approach employed here, the contribution from the $J^P=5/2^-$ state eventually surpasses that of the $J^P=3/2^-$ state, albeit at a higher energy scale. In our calculation, this crossover occurs at a center-of-mass energy of approximately $W \sim 25$ GeV.

From a partial wave perspective, $J^P=3/2^-$ states couple to $\rho p$ and $\phi p$ via $s$-wave, whereas $J^P=5/2^-$ states couple via $d$-wave. This implies that at sufficiently high energies, states with higher orbital angular momenta are more likely to be observed than those with lower partial waves.

\subsubsection{Spin Density Matrix}

We now turn to polarization observables within the hadronic molecular picture, focusing on the spin density matrix of the produced $\phi$ meson. Building on previous calculations, we consider not only the ground-state proton but also molecular states with $J^P=3/2^-$ and $5/2^-$. Additionally, following the treatment of the $N(1535)1/2^-$ molecular state in Ref.\cite{Xie:2007qt}, we include the $J^P=1/2^-$ case. It is worth noting that in Ref.\cite{Xie:2007qt}, although both $\pi$ and $\rho$ meson exchanges were considered (as both involve $s$-wave interactions with $N(1535)1/2^-$), pion exchange was found to dominate. In contrast, for the $J^P=3/2^-$ states introduced here, $\rho$ meson exchange provides the only $s$-wave interaction and should dominate. To enable a direct comparison of spin-dependent polarization effects, we restrict our calculation for $N(1535)1/2^-$ to its $\rho$ meson exchange component, following Ref.\cite{Xie:2007qt}.

\begin{table}[H]
    \caption{Spin density matrix element $\rho_{00}$ for different spin-parity at the energy point which proton beam momentum corresponding to $p_{lab}=3.67$ GeV$/c$.}
    \belowrulesep=0pt
\aboverulesep=0pt
\renewcommand{\arraystretch}{1.5}
    \centering
    \begin{tabular}{c|c|c|c}
   ~$J^P$~ & $1/2^+(\kappa_\phi=-0.5)$& $1/2^+(\kappa_\phi=-4.0)$  & ~~~~~~~~$1/2^-$~~~~~~~~ \\\hline
   $\rho_{00}$ & $0.83$ & $0.38$ & $0.35$ \\\hline\hline
   $J^P$& ~$3/2^-(g_1)$~ & ~$3/2^-(g_2)$~& ~$3/2^-(g_3)$~\\\hline
   $\rho_{00}$ & $0.36$ & $0.35$ & $0.38$ \\\hline\hline
   $J^P$ & ~$5/2^-(g_1)$~ & ~$5/2^-(g_2)$~& ~$5/2^-(g_3)$~\\\hline
   $\rho_{00}$ & $0.20$ & $0.35$ & $0.43$ 
    \end{tabular}
    \label{tab_rho00}
\end{table}

In total, four spin-parity cases are considered: three from the hadronic molecular picture ($J^P=1/2^-$, $3/2^-$, and $5/2^-$) and one from the baryon ground state (the proton with $J^P=1/2^+$).

Using the experimental value $\rho_{00}=0.23\pm0.04$ extracted by the DISTO collaboration at a proton beam momentum of $p_{\text{lab}}=3.67$ GeV$/c$~\cite{DISTO:2000dfs}, we have computed the spin density matrix element $\rho_{00}$ for different spin-parity assignments at the same energy. The results are summarized in Tab.~\ref{tab_rho00}.

As shown in Tab.~\ref{tab_rho00}, although the tensor term ($\kappa_{\phi}$ term) of the $pp\phi$ interaction partially reduces the value of the spin density matrix element $\rho_{00}$, the result remains significantly higher than the experimental measurement. Similarly, the calculated $\rho_{00}$ values for both $J^P = 1/2^-$ and $J^P = 3/2^-$ states are nearly identical at approximately 0.36, and also substantially exceed the experimentally extracted value.

\begin{figure}[H]
    \vglue 6pt
    \centering
    \includegraphics[width=0.35\textwidth]{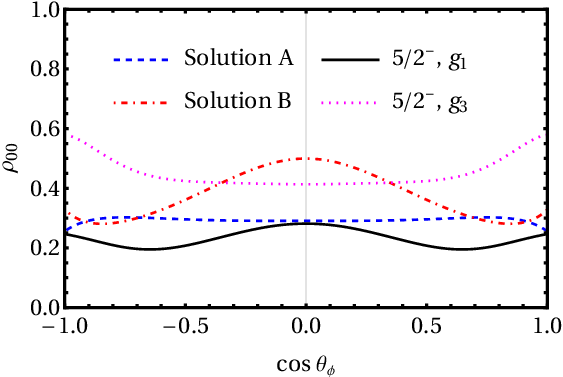}
    \caption{Spin density matrix element $\rho_{00}$ with the angle of outgoing $\phi$ meson $\theta_\phi$ in the center-of-mass frame at the center-of-mass energy $W=4.38$ GeV. The blue dashed line and the red dot-dashed line correspond to the solutions A and B obtained by our model, while the black solid line corresponds to the case of a $J^P = 5/2^-$ state with its $g_1$ term providing the dominant contribution.}
    \label{fig_rho}
\end{figure}

According to our results, only the $g_1$ term of the $J^P = 5/2^-$ state yields a $\rho_{00}$ value that falls within the experimental error band. This suggests that, within the hadronic molecular picture, a $J^P = 5/2^-$ state with a dominant $g_1$ term contribution may be required near this energy to explain the experimentally observed spin density matrix element $\rho_{00} = 0.23 \pm 0.04$.

For the center-of-mass energy $W=4.38$ GeV, corresponding to a proton beam energy of $E_p=9.3$ GeV, we have computed the spin density matrix element $\rho_{00}$ for four distinct scenarios. The first and second scenarios correspond to Solutions A and B of our model, which include the molecular states $N(2080)3/2^-$ and $N(2270)3/2^-$. The third scenario involves the aforementioned $J^P=5/2^-$ state with dominant $g_1$ term contribution, which reproduces the existing experimental data near threshold. The fourth scenario features the $J^P=5/2^-$ state with $g_3$ term dominance, whose partial wave composition differs most significantly from the $J^P=3/2^-$ cases, as shown in Figs.\ref{fig_higher_spin} and~\ref{fig_higher_spin2}.

As shown in Fig.\ref{fig_rho}, the blue dashed, red dot-dashed, black solid, and magenta dotted lines correspond to Solution A, Solution B, the $g_1$-dominant $J^P=5/2^-$ case, and the $g_3$-dominant $J^P=5/2^-$ case, respectively. The four curves exhibit distinct line shapes: nearly flat, centrally peaked with lower ends, symmetrically undulating, and centrally suppressed with enhanced ends.

\begin{figure}[H]
    \vglue 6pt
    \centering
    \includegraphics[width=0.35\textwidth]{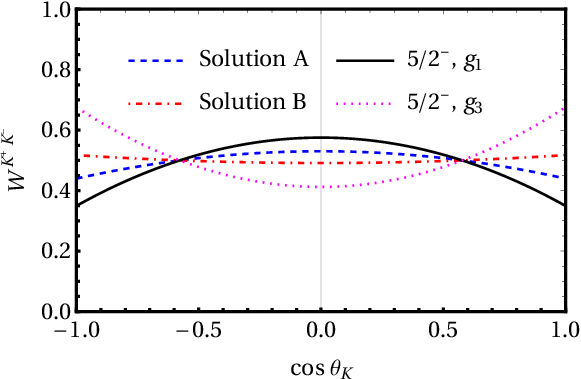}
    \caption{The distribution $\mathcal{W}^{K^+K^-}$ of $K^+K^-$ particles in the rest frame of $\phi$ meson, where $\theta_K$ is the decay angle. Notation is the same as in Fig.\ref{fig_rho}.}
    \label{fig_KK}
\end{figure}

For further discrimination, we present in Fig.\ref{fig_KK} the angular distribution $\mathcal{W}^{K^+K^-}$ of $K^+K^-$ pairs in the $\phi$ meson rest frame. Here, the red dot-dashed line (Solution B) approaches a nearly straight distribution, while the blue dashed line (Solution A) and the black solid line ($J^P=5/2^-$ with $g_1$ term dominance) exhibit progressively stronger curvature. In contrast, the magenta dotted line ($J^P=5/2^-$ with $g_3$ term dominance) shows an opposite curvature pattern.

Based on these polarization observable predictions, we anticipate that future experiments will clearly distinguish the corresponding polarization properties, thereby providing further constraints on theoretical models.

Beyond the discussions above, a recent measurement from STAR Collaboration in heavy-ion collisions has revealed a striking discrepancy with theoretical predictions as the global spin alignment which is $\rho_{00}$ for $\phi$ is unexpectedly large~\cite{STAR:2022fan}, and the observed spin-alignment pattern and magnitude for the $\phi$ cannot be explained by conventional mechanisms, while a model with a connection to strong force fields~\cite{Liang:2004ph,Sheng:2019kmk,Sheng:2020ghv}.

In their report, the $\rho_{00}$ for $\phi$ mesons, averaged over $\sqrt{s_{NN}}$ between $11.5$ and
$62$ GeV is $0.3512 \pm 0.0017 (stat.) \pm 0.0017 (syst.)$ and the results indicate that the $\phi$-meson $\rho_{00}$ is above $1/3 $ with a significance of $7.4~\sigma$. Theorists have also put forward explanations~\cite{Lv:2024uev}.

\begin{table}[H]
    \caption{Spin density matrix element $\rho_{00}$ for solutions A and B,and the case of $J^P = 5/2^-$ state with its $g_3$ term dominance.}
    \belowrulesep=0pt
\aboverulesep=0pt
\renewcommand{\arraystretch}{1.5}
    \centering
    \begin{tabular}{c|c|c|c}
   ~$\sqrt{s_{NN}}$~ & ~Solution A~ & ~Solution B \cmmnt{& $5/2^-~(g_1)$} & ~$5/2^-~(g_3)$~ \\\hline
   $11.5$ GeV & $0.2630$ & $0.3254$ \cmmnt{& $0.3528$} & $0.4775$ \\
   $19.6$ GeV & $0.2662$ & $0.3213$ \cmmnt{& $0.2866$} & $0.5082$ \\
   $27.0$ GeV & $0.2735$ & $0.3270$ \cmmnt{& $0.3571$} & $0.5324$ \\
   $39.0$ GeV & $0.2573$ & $0.3098$ \cmmnt{& $0.4453$} & $0.5387$ \\
   $62.4$ GeV & $0.3141$ & $0.3333$ \cmmnt{& $0.1613$} & $0.5376$ 
    \end{tabular}
    \label{tab_438}
\end{table}

The essence of heavy-ion collisions can be understood as interactions between densely clustered nucleon aggregates, where individual component behaviors influence the collective system dynamics. Within this framework, predictions from proton-proton scattering remain highly relevant for interpreting heavy-ion phenomena. Following this approach, we calculate the spin density matrix element $\rho_{00}$ for Solutions A and B, as well as for the $J^P=5/2^-$ state with dominant $g_3$ term which is expected to contribute most significantly based on Fig.~\ref{fig_higher_spin}. Calculations are performed at $\sqrt{s_{NN}}=11.5$, 19.6, 27.0, 39.0, and 62.4 GeV, corresponding to energy points reported by the STAR Collaboration, with results listed in Tab.~\ref{tab_438}.

Our calculations demonstrate that the observed anomalous $\rho_{00}$ enhancement can be explained by internal production mechanisms involving resonance states. As shown in Tab.~\ref{tab_438}, the three cases exhibit distinct patterns: Solution A yields $\rho_{00} ~\textless~ 1/3$, Solution B gives $\rho_{00} \approx 1/3$, while the $J^P=5/2^-$ state with $g_3$ dominance produces $\rho_{00} ~\textgreater~ 1/3$. Considering that these excited resonances may mix collectively, this opens the possibility of explaining the global $\phi$ meson spin alignment through resonant states—a mechanism we term the Deeply-Internal Resonances Excitation Mechanism (DiREM).

\begin{figure}[H]
    \vglue 6pt
    \centering
    \includegraphics[width=0.36\textwidth]{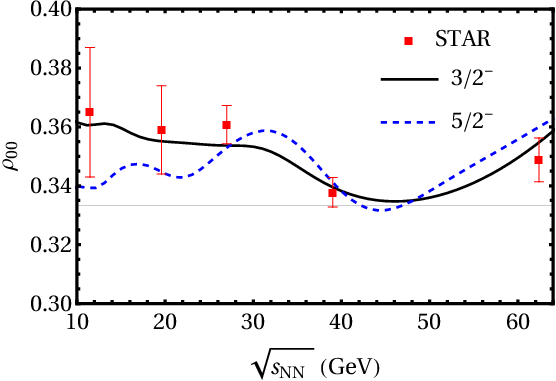}
    \caption{A simple fit by using DiREM with $N(2270)$, where the red points represent the data from the STAR collaboration, and the black solid line and the blue dashed line correspond to our fit of case $J^P = 3/2^-$ and $5/2^-$, respectively, while the gray grid line represents $\rho_{00}=1/3$.}
    \label{fig_STAR}
\end{figure}

To further demonstrate the viability of DiREM, we performed a fit to the STAR dataset by introducing a single $N(2270)$ resonance with spin-parity $J^P = 3/2^-$ or $5/2^-$ within the molecular picture. By adjusting the coupling constant ratios in Eqs.~\eqref{eq_3half} or \eqref{eq_5h}, we achieved a satisfactory description of the high-precision experimental data from proton-proton scattering, with $\chi^2 = 0.69$ or $2.16$, respectively. The corresponding fitting parameters are listed in Tab.~\ref{tab_STAR}. As shown in Fig.~\ref{fig_STAR}, the red points represent the STAR collaboration data, while the black solid line and the blue dashed line correspond to our fit of case $J^P = 3/2^-$ and $5/2^-$, respectively.

\begin{table}[H]
    \caption{Fitting parameters of STAR Collaboration measurement of global spin alignment by using DiREM with $N(2270)$. } 
    \belowrulesep=0pt
    \aboverulesep=0pt
    \renewcommand{\arraystretch}{1.5}
    \centering
    \begin{tabular}{c|cc}
   ~~~$J^P$~~~ & ~~~Parameter &~~~~~~~~~~~~~ Value $\pm$ Error \\\hline
       $3/2^-$ & $g_2/g_1$ &~~~~~~~~~~~  $-0.19\pm0.16$ \\
      $(\chi^2=0.69)$ & $g_3/g_1$ &~~~~~~~~~~~~~  $1.24\pm0.04$\\\hline
         $5/2^-$ & $g_2/g_1$ &~~~~~~~~~~~  $-1.24\pm0.16$ \\
      $(\chi^2=2.16)$ & $g_3/g_1$ &~~~~~~~~~~~~~  $3.45\pm0.70$\\\hline
    \end{tabular}
    \label{tab_STAR}
\end{table}

It is noteworthy that we applied the same fitting procedure to both spin-parity assignments. For the $3/2^-$ case, the large error of the $g_2/g_1$ indicates that the experimental data can be fully described using only the parameter $g_3/g_1$. This finding strongly supports the consistency between our proposed DiREM and existing experimental observations.

Assuming that a similar DiREM operates in heavy-ion collisions, both a pure DiREM scenario and a hybrid scenario—where only a fraction of colliding particles excite resonance states—could provide viable explanations for the $\phi$ meson spin alignment results reported by the STAR Collaboration.

\subsubsection{Perspective of Experiments}

To facilitate experimental measurements at the center-of-mass energy $W=4.38$ GeV (corresponding to a proton beam energy of $E_p=9.3$ GeV), we employed Monte Carlo simulations to reconstruct the four-momenta of final-state particles. We further computed the laboratory-frame angular distributions and momentum spectra for the final-state proton and kaon from the process $pp\rightarrow ppK^+K^-$.

\begin{figure}[H]
    \vglue 6pt
    \centering
    \includegraphics[width=0.47\textwidth]{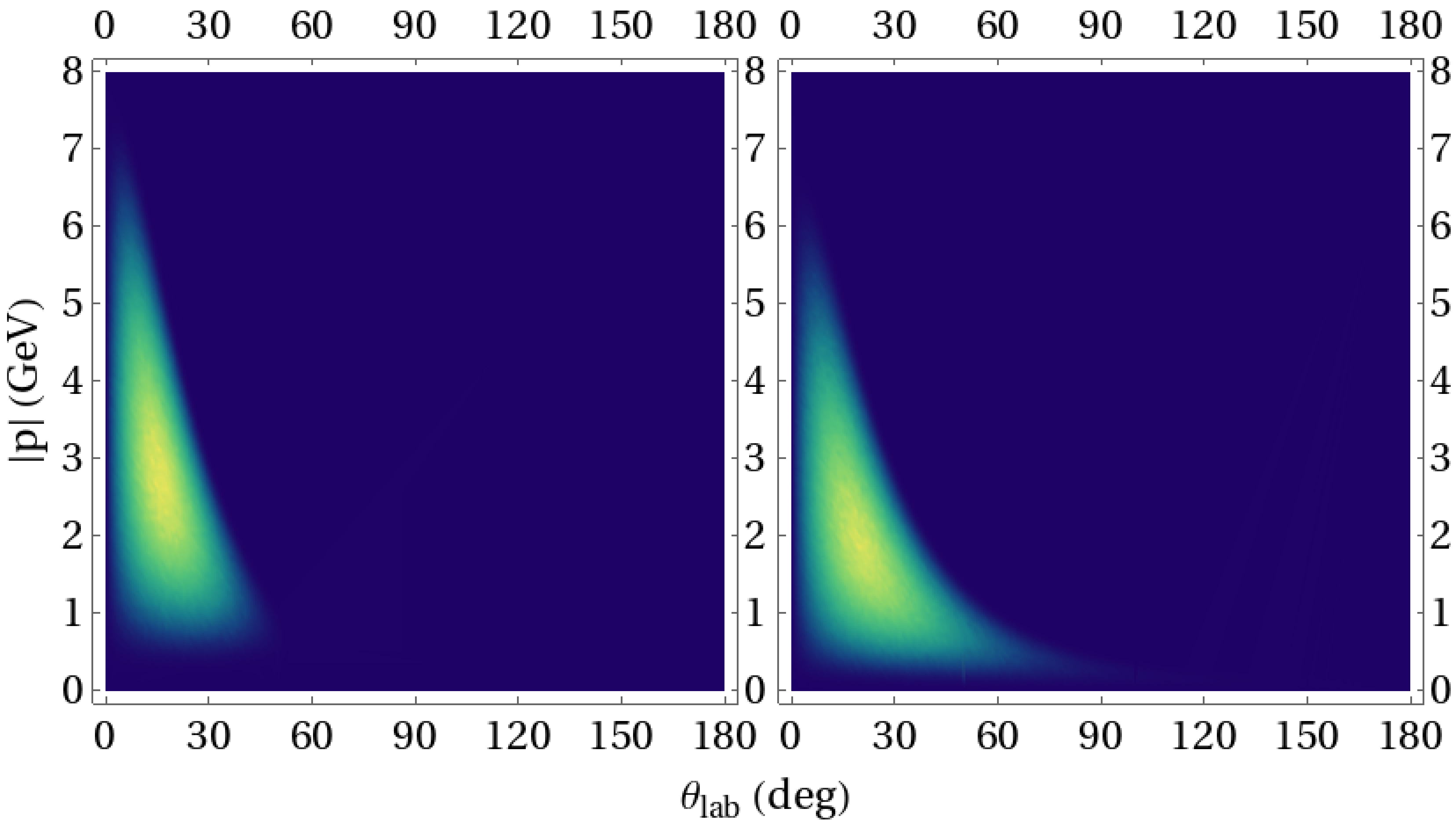}
    \caption{Distribution of the final-state proton(left panel) and kaon(right panel) at $W=4.38$ GeV in $pp\rightarrow ppK^+K^-$. Where the vertical axis represents the magnitude of the particle momentum, and the horizontal axis denotes the outgoing angle relative to the beam axis in angular units, while lighter shades indicate higher distribution density, and darker shades correspond to lower density.}
    \label{fig_438}
\end{figure}

As shown in Fig.~\ref{fig_438}, most protons are concentrated within a momentum range of $0.6$--$7.0$ GeV and scattering angles below $45^\circ$, while most kaons populate the momentum region of $0.3$--$6.0$ GeV with scattering angles up to $75^\circ$.

These results are expected to provide valuable guidance for the design and optimization of future detector systems in experimental facilities.

\subsection{$pp\rightarrow ppJ/\psi$}

\subsubsection{Cross Section}

\begin{figure}[H]
    \vglue 6pt
    \centering
    \includegraphics[width=0.36\textwidth]{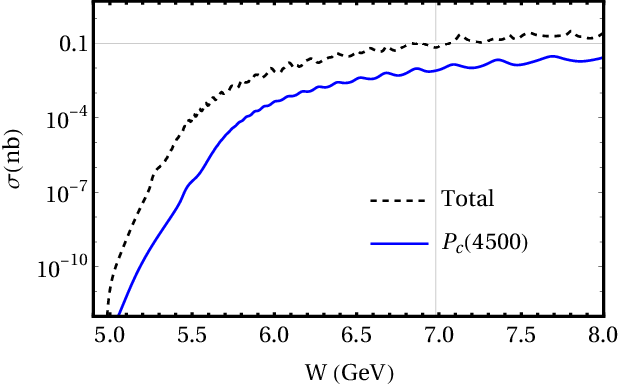}
    \caption{Total cross section of $pp\rightarrow ppJ/\psi$ by exchanging $\rho$ meson and $P_c$ states. Where the black dashed line and blue solid line correspond to the total production cross section and the individual contribution from $P_c(4500)$.}
    \label{fig_jpsi}
\end{figure}

We present a concise analysis and corresponding theoretical predictions for $J/\psi$ production in proton-proton scattering. The black dashed line in Fig.~\ref{fig_jpsi} shows the calculated cross section for $pp\rightarrow pp J/\psi$. By introducing the $P_c$ states listed in Tab.~\ref{tab_Pc}—interpreted as hidden-charm hadronic molecular states—the total $J/\psi$ production cross section reaches approximately $0.1$~nb.

The apparent non-smooth behavior of the predicted curve arises from the individual contributions of $P_c$ states. As an example, the blue solid line in Fig.~\ref{fig_jpsi} shows the contribution from the broad $P_c(4500)$ state. The on-shell production of narrow $P_c$ resonances induces oscillatory structures in the cross section, with this effect becoming more pronounced for smaller decay widths. Moreover, the inclusion of six distinct $P_c$ states and non-trivial interference among their amplitudes collectively account for the observed irregularities.

\begin{figure}[H]
    \vglue 6pt
    \centering
    \includegraphics[width=0.35\textwidth]{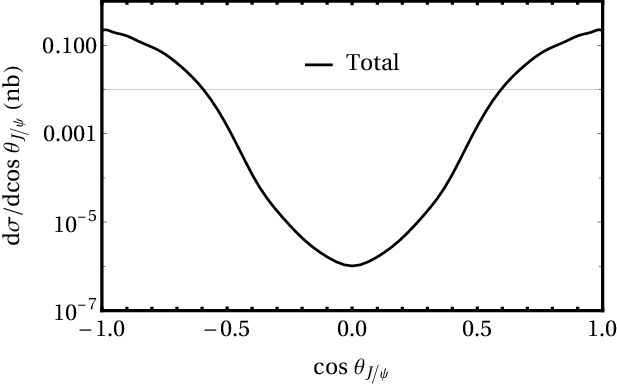}
    \caption{Differential cross section corresponding to the outgoing angle of $J/\psi$ meson $\theta_{J/\psi}$ at center-of-mass energy $W = 6.98$ GeV.}
    \label{fig_jdif}
\end{figure}

To analyze specific physical observables at well-defined energies, we perform calculations at $W=6.98$ GeV, corresponding to a proton beam energy of $E_p=25$ GeV as proposed for the CNUF. As shown in Fig.~\ref{fig_jdif}, $J/\psi$ mesons are predominantly produced near $\cos\theta_{J/\psi} \approx \pm1$, indicating strong forward-backward peaking along the beam direction in the center-of-mass frame.

\subsubsection{Spin Density Matrix}

\begin{figure}[H]
    \vglue 6pt
    \centering
    \includegraphics[width=0.34\textwidth]{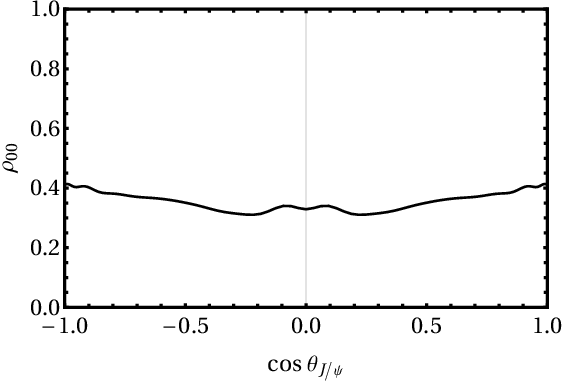}
    \caption{Spin density matrix element $\rho_{00}$ corresponding to the outgoing angle of $J/\psi$ meson $\theta_{J/\psi}$ at center-of-mass energy $W = 6.98$ GeV.}
    \label{fig_rhoj}
\end{figure}

We now extend our discussion to spin polarization observables at the center-of-mass energy $W=6.98$ GeV. Fig.~\ref{fig_rhoj} shows the spin density matrix element $\rho_{00}$ as a function of the $J/\psi$ production angle $\theta_{J/\psi}$ at this energy.

\begin{figure}[H]
    \vglue 6pt
    \centering
    \includegraphics[width=0.34\textwidth]{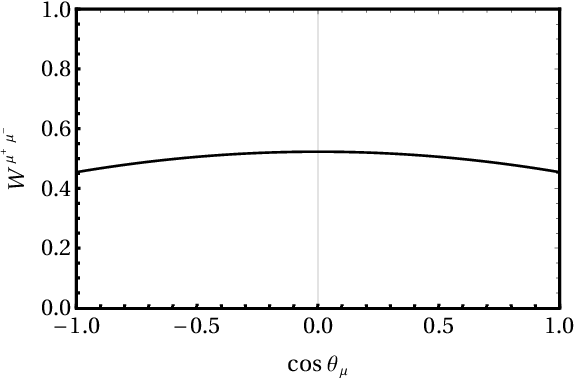}
    \caption{The distribution $\mathcal{W}^{\mu^+\mu^-}$ of $\mu^+\mu^-$ particles in the rest frame of $J/\psi$ meson, where $\theta_\mu$ is the decay angle.}
    \label{fig_mumu}
\end{figure}

Additionally, Fig.~\ref{fig_mumu} presents the angular distribution $\mathcal{W}^{\mu^+\mu^-}$ of $\mu^+\mu^-$ pairs in the $J/\psi$ rest frame, which exhibits only mild curvature.

\subsubsection{Perspective of Experiments}

To facilitate experimental measurements at the center-of-mass energy $W=6.98$ GeV (corresponding to a proton beam energy of $E_p=25$ GeV), we employed Monte Carlo simulations to reconstruct the four-momenta of final-state particles. We further computed the laboratory-frame angular distributions and momentum spectra for the final-state proton and muon from the process $pp\rightarrow pp\mu^+\mu^-$.

\begin{figure}[H]
    \vglue 6pt
    \centering
    \includegraphics[width=0.48\textwidth]{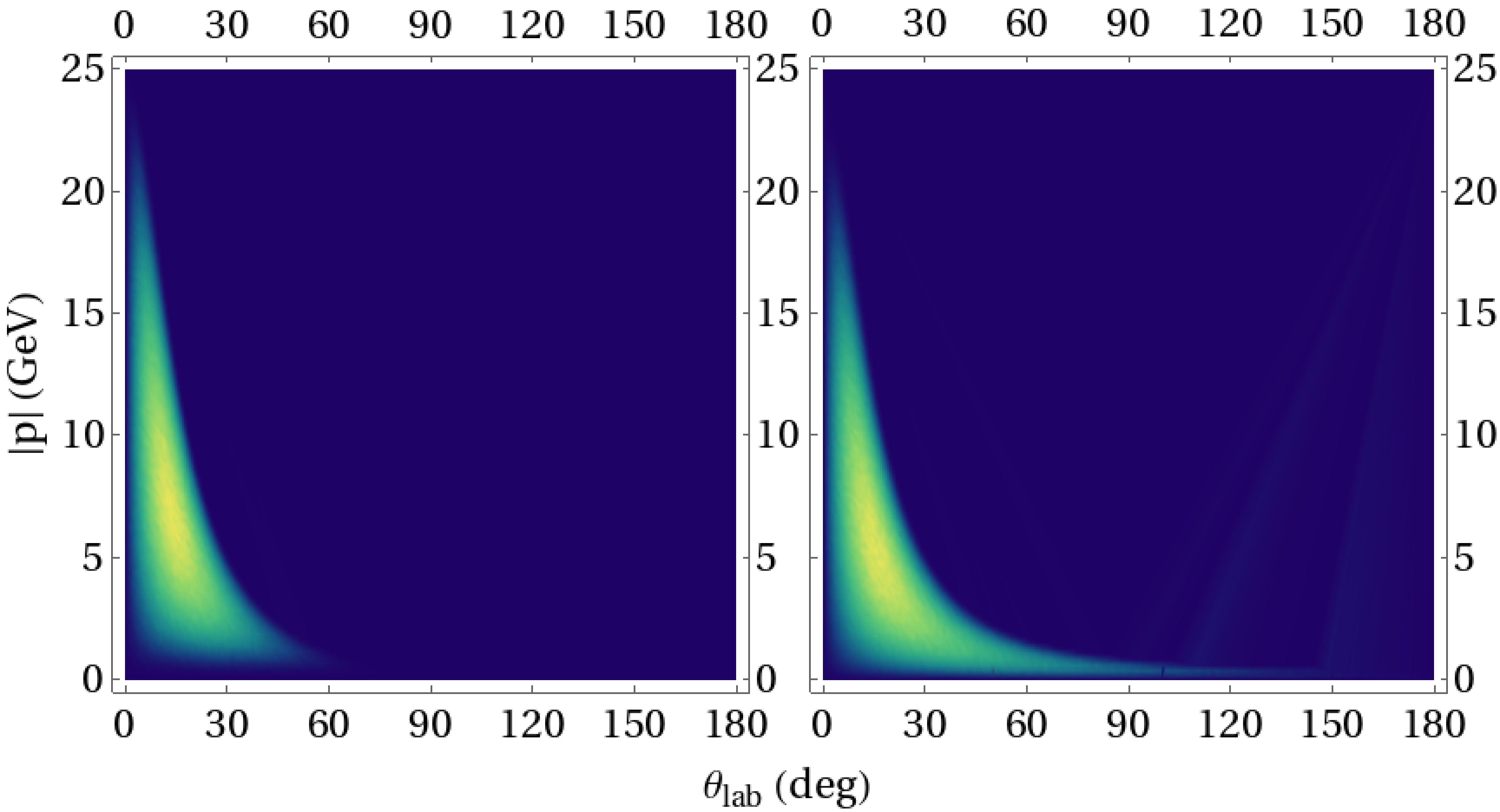}
    \caption{Distribution of the final-state proton(left panel) and muon(right panel) at $W=6.98$ GeV in $pp\rightarrow pp\mu^+\mu^-$. Notation is the same as Fig.~\ref{fig_438}}
    \label{fig_698}
\end{figure}

As shown in Fig.~\ref{fig_698}, most protons are concentrated within a momentum range of $1$--$22$ GeV and scattering angles below $60^\circ$. In contrast, muons predominantly populate the momentum region below $20$ GeV but extend to much larger scattering angles, up to $135^\circ$.

A substantial yield of low-energy muons at large angles constitutes the main difference compared to the $pp\rightarrow pp\phi$ case. These results are expected to provide valuable guidance for the design and optimization of detector systems in future experimental facilities.

\section{Summary and Conclusions} \label{sec:summary}

In this article, we investigate $\phi$ and $J/\psi$ meson production in proton-proton scattering within the hadronic molecular picture.

For $pp\rightarrow pp\phi$, we estimate the $\phi$ production cross section by introducing the hadronic molecular states $N(2080)3/2^-$ and $N(2270)3/2^-$. Our calculations indicate that the cross section reaches the order of $10~\mu$b within the energy range accessible at the HIAF facility. We also predict the angular distribution of the $\phi$ meson at a center-of-mass energy of $W=4.38$ GeV.

Further analysis of the spin density matrix suggests that explaining existing experimental data likely requires the inclusion of a $J^P=5/2^-$ state near the production threshold. We provide model predictions for the $\phi$ emission angle dependence of $\rho_{00}$ in the center-of-mass frame, as well as the kaon angular distribution from $\phi \rightarrow K^+K^-$ decay in the $\phi$ rest frame.

Inspired by the STAR Collaboration results, we propose a Deeply-Internal Resonances Excitation Mechanism (DiREM). Analysis of the $\phi$ spin density matrix shows that both a pure resonance production scenario and a hybrid approach—where only a fraction of colliding particles produce resonances—offer a novel explanation for the observed large global spin alignment exceeding $1/3$.

From the experimental perspective, we perform Monte Carlo simulations of the four-body final state in $pp\rightarrow ppK^+K^-$. We find that most protons are concentrated within a momentum range of $0.6$--$7.0$ GeV and scattering angles below $45^\circ$, while kaons primarily populate the $0.3$--$6.0$ GeV momentum range with scattering angles up to $75^\circ$.

For $pp\rightarrow ppJ/\psi$, we estimate the production cross section by introducing six $P_c$ states within the hadronic molecular picture. The calculated $J/\psi$ cross section reaches approximately $0.1$~nb within the proposed CNUF energy range. We also predict the $J/\psi$ angular distribution at $W=6.98$ GeV.

Additionally, we provide the $J/\psi$ emission angle dependence of $\rho_{00}$ in the center-of-mass frame, along with the muon angular distribution from $J/\psi \rightarrow \mu^+\mu^-$ decay in the $J/\psi$ rest frame.

Monte Carlo simulations of $pp\rightarrow pp\mu^+\mu^-$ reveal that protons mainly occupy the $1$--$22$ GeV momentum range with scattering angles below $60^\circ$, while muons exhibit a broad angular distribution up to $135^\circ$, with a substantial yield of low-energy muons at large angles.

We hope these calculations provide practical guidance for future experiments at HIAF and CNUF, enabling more precise measurements and thereby imposing stronger constraints on theoretical models.

\begin{acknowledgments}
The authors appreciate the meaningful discussion with Shu-Ming Wu, Jia-Jun Wu, and Feng-Kun Guo. 
\end{acknowledgments}

\bibliographystyle{unsrt}

\begin{thebibliography}{99}
%
\bibitem{Hartmann:2005wj}
M.~Hartmann, Y.~Maeda, I.~Keshelashvili, H.~R.~Koch, S.~Mikirtychyants, S.~Barsov, W.~Borgs, M.~Buescher, V.~Hejny and V.~Kleber, \textit{et al.}

\bibitem{Hartmann:2006zc}
M.~Hartmann, Y.~Maeda, I.~Keshelashvilli, H.~R.~Koch, S.~Mikirtytchiants, S.~Barsov, W.~Borgs, M.~Buscher, V.~I.~Dimitrov and S.~Dymov, \textit{et al.}
Phys. Rev. Lett. \textbf{96}, 242301 (2006)
[erratum: Phys. Rev. Lett. \textbf{97}, 029901 (2006)]

\bibitem{DISTO:1998hzu}
F.~Balestra \textit{et al.} [DISTO],
Phys. Rev. Lett. \textbf{81}, 4572-4575 (1998)

\bibitem{DISTO:2000dfs}
F.~Balestra \textit{et al.} [DISTO],
Phys. Rev. C \textbf{63}, 024004 (2001)

\bibitem{Xie:2007qt}
J.~J.~Xie, B.~S.~Zou and H.~C.~Chiang,
Phys. Rev. C \textbf{77}, 015206 (2008)

\bibitem{Tsushima:2003fs}
K.~Tsushima and K.~Nakayama,
Phys. Rev. C \textbf{68}, 034612 (2003)
doi:10.1103/PhysRevC.68.034612
[arXiv:nucl-th/0304017 [nucl-th]].

\bibitem{Yang:2013yeb}
J.~C.~Yang, J.~W.~Xia, G.~Q.~Xiao, H.~S.~Xu, H.~W.~Zhao, X.~H.~Zhou, X.~W.~Ma, Y.~He, L.~Z.~Ma and D.~Q.~Gao, \textit{et al.}
Nucl. Instrum. Meth. B \textbf{317}, 263-265 (2013)

\bibitem{Zhou:2022pxl}
X.~Zhou \textit{et al.} [HIAF project Team],
AAPPS Bull. \textbf{32}, no.1, 35 (2022)

\bibitem{An:2025lws}
F.~An, D.~Bai, S.~Chen, X.~Chen, H.~Duyang, L.~Gao, S.~F.~Ge, J.~He, J.~Huang and Z.~Huang, \textit{et al.}
[arXiv:2504.21050 [hep-ph]].

\bibitem{Aaij:2015tga}
R. Aaij {\it et al.} (LHCb Collaboration), Phys. Rev. Lett. {\bf 115}, 072001 (2015).
%
\bibitem{Aaij:2019}
R. Aaij {\it et al.} (LHCb Collaboration), Phys. Rev. Lett. {\bf 122}, 222001 (2019).
%

\bibitem{Wu:2010jy}
J.~J.~Wu, R.~Molina, E.~Oset and B.~S.~Zou,
Phys. Rev. Lett. \textbf{105}, 232001 (2010)

\bibitem{Wu:2010vk}
J.~J.~Wu, R.~Molina, E.~Oset and B.~S.~Zou,
Phys. Rev. C \textbf{84}, 015202 (2011)

\bibitem{Wang:2011rga}
W.~L.~Wang, F.~Huang, Z.~Y.~Zhang and B.~S.~Zou,
Phys. Rev. C \textbf{84}, 015203 (2011)

\bibitem{Wu:2012md}
J.~J.~Wu, T.~S.~H.~Lee and B.~S.~Zou,
Phys. Rev. C \textbf{85}, 044002 (2012)

\bibitem{Lin:2017mtz}
Y.~H.~Lin, C.~W.~Shen, F.~K.~Guo and B.~S.~Zou,
Phys. Rev. D \textbf{95}, no.11, 114017 (2017)

\bibitem{Shen:2017oyw}
C.~W.~Shen and Y.~H.~Lin,
Phys. Part. Nucl. Lett. \textbf{15}, no.4, 402-405 (2018)

\bibitem{Lin:2019qiv}
Y.~H.~Lin and B.~S.~Zou,
Phys. Rev. D \textbf{100}, no.5, 056005 (2019)

\bibitem{Duan:2024hby}
M.~X.~Duan, C.~Gong, L.~Qiu and Q.~Zhao,
[arXiv:2409.10364 [hep-ph]].

\bibitem{Lin:2018kcc}
Y.~H.~Lin, C.~W.~Shen and B.~S.~Zou,
Nucl. Phys. A \textbf{980}, 21-31 (2018)

\bibitem{Ben:2024qeg}
D.~Ben and S.~M.~Wu,
Phys. Rev. C \textbf{112}, no.1, 015203 (2025)

\bibitem{Wu:2023ywu}
S.~M.~Wu, F.~Wang and B.~S.~Zou,
Phys. Rev. C \textbf{108}, no.4, 045201 (2023)

\bibitem{Koch:1985bp}
R.~Koch,
Z. Phys. C \textbf{29}, 597 (1985)

\bibitem{Ronchen:2012eg}
D.~Ronchen, M.~Doring, F.~Huang, H.~Haberzettl, J.~Haidenbauer, C.~Hanhart, S.~Krewald, U.~G.~Meissner and K.~Nakayama,
Eur. Phys. J. A \textbf{49}, 44 (2013)

\bibitem{ParticleDataGroup:2024cfk}
S.~Navas \textit{et al.} [Particle Data Group],
Phys. Rev. D \textbf{110}, no.3, 030001 (2024)

\bibitem{Xie:2022hhv}
J.~M.~Xie, X.~Z.~Ling, M.~Z.~Liu and L.~S.~Geng,
Eur. Phys. J. C \textbf{82}, no.11, 1061 (2022)
doi:10.1140/epjc/s10052-022-11026-0
[arXiv:2204.12356 [hep-ph]].

\bibitem{Guidal:1997hy}
M.~Guidal, J.~M.~Laget and M.~Vanderhaeghen,
Nucl. Phys. A \textbf{627}, 645-678 (1997)

\bibitem{Titov:1997kt}
A.~I.~Titov, B.~Kampfer and V.~V.~Shklyar,
Phys. Rev. C \textbf{59}, 999-1008 (1999)

\bibitem{STAR:2022fan}
M.~S.~Abdallah \textit{et al.} [STAR],
Nature \textbf{614}, no.7947, 244-248 (2023)

\bibitem{Liang:2004ph}
Z.~T.~Liang and X.~N.~Wang,
Phys. Rev. Lett. \textbf{94}, 102301 (2005)
[erratum: Phys. Rev. Lett. \textbf{96}, 039901 (2006)]

\bibitem{Sheng:2019kmk}
X.~L.~Sheng, L.~Oliva and Q.~Wang,
Phys. Rev. D \textbf{101}, no.9, 096005 (2020)
[erratum: Phys. Rev. D \textbf{105}, no.9, 099903 (2022)]

\bibitem{Sheng:2020ghv}
X.~L.~Sheng, Q.~Wang and X.~N.~Wang,
Phys. Rev. D \textbf{102}, no.5, 056013 (2020)

\bibitem{Lv:2024uev}
J.~P.~Lv, Z.~H.~Yu, Z.~T.~Liang, Q.~Wang and X.~N.~Wang,
Phys. Rev. D \textbf{109}, no.11, 114003 (2024)









\end{thebibliography}

\end{document}